\documentclass[conference]{IEEEtran}
\IEEEoverridecommandlockouts

\usepackage{cite}
\usepackage{url}
\usepackage{amsmath,amssymb,amsfonts}
\usepackage{algorithmic}
\usepackage{multirow}
\usepackage{enumerate}
\usepackage{footmisc}
\usepackage{siunitx}
\usepackage{hyperref} 
\usepackage{array,framed}
\usepackage{booktabs}
\usepackage{etoolbox}
\usepackage{listings}  
\usepackage{xcolor}    
\usepackage{subfigure}
\usepackage{tcolorbox}
\tcbuselibrary{breakable}

\definecolor{LightGray}{RGB}{248, 248, 248}

\lstdefinestyle{customPython}{
    backgroundcolor=\color{LightGray},   
    basicstyle=\footnotesize\ttfamily,  
    breaklines=true,                     
    frame=tb,                            
    language=Python,                     
    numbers=left,                        
    numbersep=5pt,                       
    numberstyle=\tiny\color{gray},       
}

\usepackage{
  tikz,
  pgfplots,
  pgfplotstable
}
\usepackage{xcolor} 
\usepackage{
  color,
  float,
  epsfig,
  wrapfig,
  graphics,
  graphicx,
}
\usepackage{algorithm2e}
\usepackage{graphicx}
\usepackage{textcomp}
\def\BibTeX{{\rm B\kern-.05em{\sc i\kern-.025em b}\kern-.08em
    T\kern-.1667em\lower.7ex\hbox{E}\kern-.125emX}}

\newcommand{\wenboreview}[1][\textcolor{black}]{#1}
\newcommand{\camera}[1][\textcolor{black}]{#1}

\usepackage[font=small,skip=2pt]{caption}

\usepackage{microtype}
\def\BibTeX{{\rm B\kern-.05em{\sc i\kern-.025em b}\kern-.08em
    T\kern-.1667em\lower.7ex\hbox{E}\kern-.125emX}}
\begin{document}

\title{An Empirical Study of Malicious Code In PyPI Ecosystem}

\author{
\IEEEauthorblockN{Wenbo Guo\IEEEauthorrefmark{1},
Zhengzi Xu\IEEEauthorrefmark{2}\IEEEauthorrefmark{3},
Chengwei Liu\IEEEauthorrefmark{2},
Cheng Huang\IEEEauthorrefmark{1},
Yong Fang\IEEEauthorrefmark{1}\IEEEauthorrefmark{3},
Yang Liu\IEEEauthorrefmark{2}
}
\text{\small honywenair@gmail.com, 
zhengzi.xu@ntu.edu.sg}
\thanks{\IEEEauthorrefmark{3} Zhengzi Xu and Yong Fang are the corresponding authors.}
\IEEEauthorblockA{\IEEEauthorrefmark{2}School of Computer Science and Engineering, Nanyang Technological University, Singapore}
\IEEEauthorblockA{\IEEEauthorrefmark{1}School of Cyber Science and Engineering, Sichuan University, China}

}

\maketitle

\begin{abstract}
PyPI provides a convenient and accessible package management platform to developers, enabling them to quickly implement specific functions and improve work efficiency. 
However, the rapid development of the PyPI ecosystem has led to a severe problem of malicious package propagation.
Malicious developers disguise malicious packages as normal, posing a significant security risk to end-users. 

\wenboreview{To this end, we conducted an empirical study to understand the characteristics and current state of the malicious code lifecycle in the PyPI ecosystem. We first built an automated data collection framework and collated a multi-source malicious code dataset containing 4,669 malicious package files. We preliminarily classified these malicious code into five categories based on malicious behaviour characteristics. Our research found that over 50\% of malicious code exhibits multiple malicious behaviours, with information stealing and command execution being particularly prevalent. In addition, we observed several novel attack vectors and anti-detection techniques. Our analysis revealed that 74.81\% of all malicious packages successfully entered end-user projects through source code installation, thereby increasing security risks. A real-world investigation showed that many reported malicious packages persist in PyPI mirror servers globally, with over 72\% remaining for an extended period after being discovered. Finally, we sketched a portrait of the malicious code lifecycle in the PyPI ecosystem, effectively reflecting the characteristics of malicious code at different stages. We also present some suggested mitigations to improve the security of the Python open-source ecosystem.}
\end{abstract}


\section{Introduction}

Python is widely used in the development of various program systems. It provides an official third-party repository (i.e. PyPI.org~\cite{PyPI.org}), containing a vast number of reusable package files to expedite project development.
Additionally, online software source code hosting service platforms like GitHub and Gitee also host a large number of reusable packages.
While this provides great convenience for users to efficiently utilize various functionality, the platform itself does not guarantee the security of the content. 
Consequently, criminals can upload malicious software libraries and package files\cite{bagmar2021know}, confusing developers and leading them to unwittingly download and import these packages into their own projects~\cite{ladisa2022sok}.
These code with malicious behaviors, such as implanting backdoors~\cite{PyPIbackdoor} and stealing sensitive system information~\cite{stealinformation}, pose significant risks to software security.
For instance, the Easyfuncsys~\cite{Easyfuncsys} package is malicious which steals the "leveldb" file stored on the user's device, extracts the token information, and sends it to a remote server. 
This software package has been downloaded over 1,045 times, posing a significant threat to the personal privacy of users.

In the PyPI ecosystem, current empirical research mainly focuses on the malware package level\cite{kaplan2021survey,alfadel2023empirical,valiev2018ecosystem}, and related issues at the source code level have not been explored in depth. However, the lack of feasible empirical studies limits our insight into the characteristics of malicious code in the ecosystem. Empirical studies can reveal the propagation patterns, influencing factors, and potential hazards of malicious code, thus providing a trustworthy basis for researchers to develop more efficient detection methods. With empirical research, it is easier to ensure that the proposed detection methods effectively address the issue of malicious code.

Moreover, for the empirical study, malicious code datasets are the cornerstone. Building a reliable and high-quality PyPI malicious package dataset is essential to profoundly investigate the properties and sources of malicious code in the PyPI ecosystem. However, there is no universally accepted and publicly available PyPI malicious package dataset. While Ohm et al.~\cite{ohm2020backstabber} established a python malicious code dataset called Backstabbers-Knife-Collection, it only contains 250 PyPI malicious packages and lacks sufficient sample features and metadata information. Therefore, researchers face difficulties developing and studying malicious code detection algorithms based on public data. It is necessary to develop comprehensive datasets to provide reliable data support.

In order to fill this research gap, this paper aims to conduct an empirical analysis of malicious code within the PyPI ecosystem. We first build an automated malicious code collection framework to collect a high-quality dataset of available malicious code through PyPI mirrors and other sources. 

Then we set four RQs to investigate the lifecycle stages of malicious code in the PyPI ecosystem. \camera{The research starts from the malicious contributing developers, and conducts in-depth analysis on the attributes (RQ1) and attack tactics (RQ2) of malicious code to reveal the intentions of attackers. Subsequently, focus on how malicious code evades detection and is distributed and propagated on the PyPI platform (RQ3). Finally, an investigation is conducted into the impact and infiltration (RQ4) of malicious code on enduser systems. These four research questions traverse the entire lifecycle, profoundly delving into the behavior and impact of malicious code at different stages, while providing comprehensive insights pertinent to software supply chain security.}

\begin{itemize}
\item \textbf{RQ1: Code Attributes.} What are the primary attributes and sources of malicious code in the PyPI ecosystem, and how do they compare to other platforms?

\item \textbf{RQ2: Attack Tactics.} How do attackers combine various attack strategies and malicious behaviors when injecting code into open-source packages in the PyPI ecosystem, and how do these tactics evolve and adapt across different platforms and objectives?

\item \textbf{RQ3: Evasion Techniques and Distribution.} How effective are existing detection tools at identifying malicious packages, which evasion methods do malicious code employ to elude these tools, and what is the impact of these methods on their distribution in real-world applications?

\item \textbf{RQ4: Impact and Infiltration.} Within the PyPI ecosystem, how has the impact of malicious packages on end-users evolved over time, which operating systems have been affected, and what methods have been employed by malicious packages to infiltrate user systems?
\end{itemize}

The main contribution of the study can be summarized in the following points.

(1). We constructed a dataset of 4,467 malicious code, covering malware samples, malicious code snippets and PyPI malware packages. Notably, the PyPI malware package subset contains 2,035 instances, \wenboreview{making it the largest publicly available PyPI malicious packages dataset.} Moreover, the dataset construction method presented in this paper can be applied to other programming languages as well.

(2). We built an automated classification framework to categorize the collected malicious code into different types. As different types of malicious code have unique triggering methods and execution modes, we extracted patterns for each type of malicious code to provide researchers with better ideas for detecting malicious code.

(3). \wenboreview{We conducted a comprehensive empirical study on malicious code in the PyPI ecosystem. We analyzed the characteristics, behaviours, propagation, and impacts of malicious code from three dimensions: malicious contributors, open-source platforms, and end users. Moreover, we present the lifecycle portrait of malicious code, which effectively reveals the characteristics of malicious code at different stages.}

(4). \wenboreview{We analysed malicious code within the PyPI ecosystem at the source code level, revealing its behavioural patterns and evolutionary characteristics. We identified several novel anti-detection techniques. We also found \wenboreview{1,791} malicious packages in PyPI mirror servers. The residual rate of malicious packages in the Tsinghua mirror is as high as 89\%. Additionally, we found three attack vectors at install, import and run time.}

\section{Background}
\subsection{Malicious Code}
Malicious code refers to purposefully crafted code segments or software embedded in programs to execute unauthorized behaviours. Attackers utilize malicious code for purposes such as information stealing and system disruption. There are various types of malicious code, each with specific functions and propagation methods. Viruses, worms, and botnets can self-replicate\cite{faruk2021malware}; however, viruses require attachment to other programs to propagate and execute. Ransomware, spyware, adware, Trojan horses, and rabbits do not self-replicate but can propagate without being attached to other programs.

\subsection{Python Package Manager}
A package manager is a software tool that automates the process of installing, upgrading, configuring, and removing computer programs for a computer's operating system in a consistent manner. It helps users to manage the dependencies and libraries required by different software packages and simplify the software installation process. \wenboreview{Package managers ensure that individual software packages operate consistently and maintain compatibility with other system components. PyPI provides users with a centralized repository for searching, installing, and publishing various Python packages. However, there is a large amount of malicious code in this repository. This paper will focus on empirical analysis of this code to reveal potential security risks.}

\section{Data Collection and Classification}

To conduct a comprehensive analysis of malicious code within the PyPI ecosystem, it is crucial to construct a high-quality and reliable dataset that covers various sources of malicious code.

\begin{figure}[t!]
  \centering
  \includegraphics[scale=0.4]{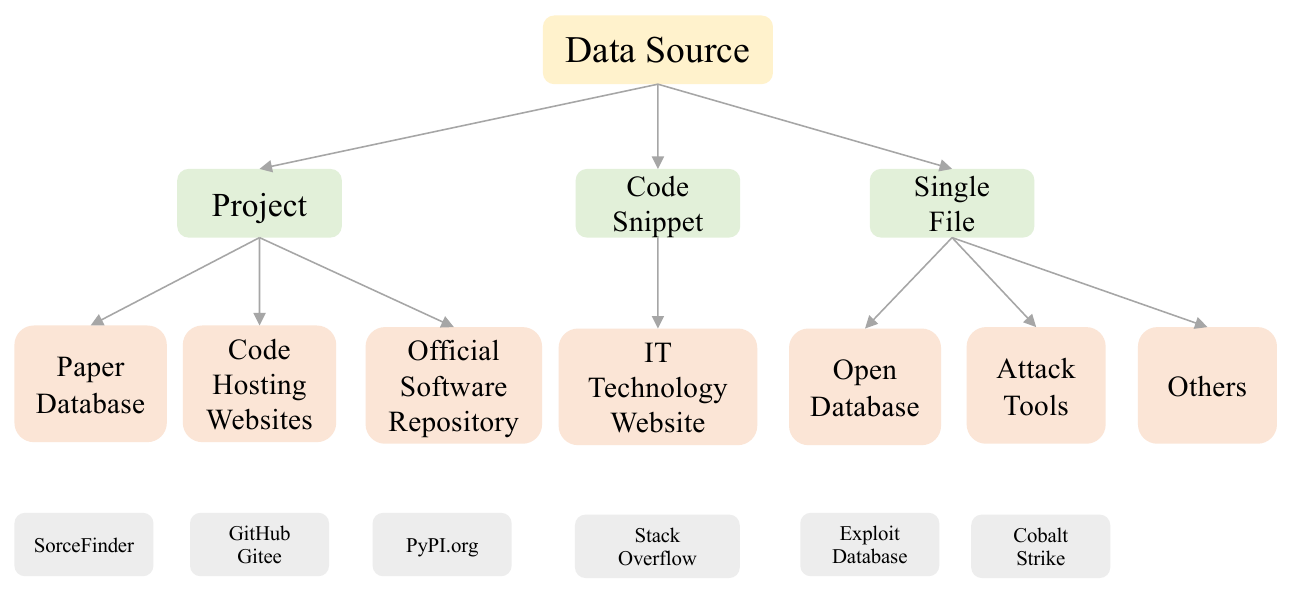}
  \caption{Data Source}
  \label{data_source}
\end{figure}

\begin{figure*}[htbp]
  \centering
  \includegraphics[scale=1.1]{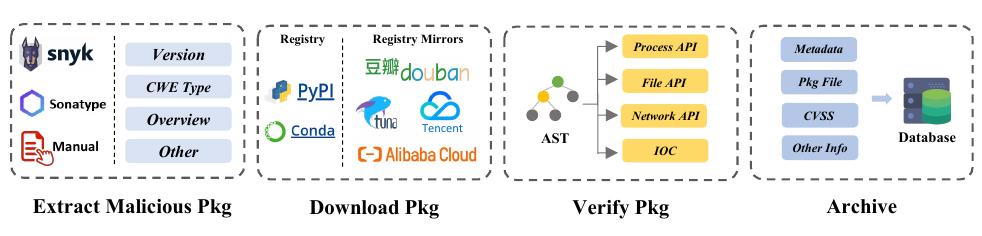}
  \caption{Malicious package collection and verification workflow}
  \label{data_collection}
\end{figure*}

To build such a dataset, we first conducted an in-depth study of relevant literature to systematically investigate the sources and construction methods of malicious code datasets. Our analysis categorized the sources of malicious code into seven classes, \wenboreview{including Paper~\cite{rokon2020sourcefinder,ohm2020backstabber,duan2020towards}, Code Hosting Websites, Official Software Repositories, IT Conversation Websites~\cite{fischer2017stack,li2017understanding, samtani2016azsecure}, Open Databases~\cite{vx-underground, exploit-db}, Attack Tools~\cite{hstechdocs,cobaltstrike} and Others, }
covering various types of malicious code, including code snippets, complete projects, and single-file forms. This classification approach helps to build a diverse dataset that better reflects the realistic distribution and attributes of malicious code. Our proposed classification method applies not only to building a Python malicious code dataset but also to other programming languages. Figure \ref{data_source} illustrates the sources and types of the dataset clearly. 

Therefore, we expect this dataset to strongly support related research and applications.

\subsection{Data Collection}

\wenboreview{We used the method to collect malicious code data for different data sources. Due to their scattered and small size, we took a manual collection approach for malicious code data from papers, open databases, attack tools, and other sources. We built a web crawler based on keyword matching for datasets from code hosting websites and IT technology websites. We selected keywords such as "malicious" and "security" to collect posts related to these keywords. Subsequently, we parsed the HTML tags of the web pages and extracted code snippets. For the malicious code data in the official software repository, we built an automated PyPI malicious package collection framework, as shown in Figure \ref{data_collection}, 
to facilitate the collection and preprocessing of malicious packages in PyPI.}

\wenboreview{Open-source intelligence sites play a crucial role in disclosing malware packages on time. We implemented an automated crawler using the Selenium library to comprehensively collect information about public PyPI malicious packages. The crawler regularly collects data from Snyk\cite{snyk} and Sonatype\cite{sonatype}  websites, including package name, version, description, CWE type, release date, and other details. After obtaining the basic information, we need to download their source code. First, we check whether the package files exist in the registry (e.g. PyPI.org and Conda). If they exist, we download them. However, for malicious packages that are not available in the package registry, we look for them in registry mirrors. Our research reveals that these mirrors do not always synchronize with the root registry and may contain packages that have been removed from the root registry. Therefore, it is still feasible to obtain source code from registry mirrors. In this process, we use Tsinghua\cite{tsinghua}, Tencent\cite{tencent}, Alibaba\cite{alibaba}, and Douban\cite{douban} registry mirrors as data sources. }

\wenboreview{Not all collected datasets contain malicious code; some instances may be false positives. We manually examined data from papers, attack tools, and other sources, finding these three categories to have relatively high data quality. We utilized a semi-automated approach to check the code files for data obtained from open databases, IT technology websites, code hosting websites, and official software repositories. Malicious packages differ from benign ones at the source code level, requiring specific APIs and IOCs (URLs, IPs, and paths) to perform malicious behaviours. Therefore, we parsed the code files into abstract syntax trees and extracted five suspicious API categories, including network APIs, file operation APIs, process APIs, encryption APIs, and execution APIs, to determine if the code is truly malicious. Additionally, we utilized VirusTotal\cite{virustotal} to verify the maliciousness of the IOCs. To address potential false positives by this method, we also conducted manual inspections of the results.}

\begin{table}
	\centering
	\renewcommand\arraystretch{1.3}
        \footnotesize
	\caption{Source and size of malicious code datasets}
        \scalebox{0.90}{
        \setlength{\tabcolsep}{1.3pt}
        \begin{tabular}{llll}
		\toprule
		\textbf{Source Type} & \textbf{Data Source} & \textbf{Number} & \textbf{Data Type} \\
		\hline
		\multirow{3}{*}{\textbf{Paper}} 
		&  SorceFinder\cite{rokon2020sourcefinder}  &  2,355  &  Project \\
            &  Maloss\cite{duan2020towards}  &  277  &  Package / Single File\\
        &  Backstabbers-Knife\cite{ohm2020backstabber}  &  250  &  Package \\
		\hline
		\multirow{2}{*}{\textbf{Code Hosting Website}} 
		&  GitHub  &  65   &  Project  \\
            &  Gitee  &  2 &    Project  \\
		\hline
		\multirow{3}{*}{\textbf{Official Software Repository}} 
		 & Snyk.io  &  1,791  &  Package \\
          & Manual  &  30  &  Package \\
          & Manual Extraction  &  253  &  Code Snippet \\
		\hline
		\multirow{2}{*}{\textbf{IT Conversation Website}} 
		&  StackOverflow  &  116  &    Code Snippet  \\
        &  Hacker forums  &   1  &    Code Snippet \\
		 \hline
		 \multirow{2}{*}{\textbf{Open Database }} 
		 &  Exploit Database  &  25   &   Single File \\
          &  VX-Underground  &  17   &   Single File / Project \\
         \hline
		 \multirow{1}{*}{\textbf{Attack Tools }} 
		 &  Cobalt Strike  &  1   &  Single File  \\
        \hline
		 \multirow{1}{*}{\textbf{Others }} 
		 &  Manual  & 1  &  Single File \\
		\bottomrule
        \multirow{1}{*}{\textbf{All }} 
		 &  Deduplication  & 4,669  &   \\
		\bottomrule
	\end{tabular}}
	\label{dataset}
\end{table}

For subsequent analysis, we manually extracted malicious code from selected PyPI packages and added these malicious code snippets to the database in file form. Finally, we constructed a multi-source malicious code dataset, and Table \ref{dataset} provides detailed statistics. The dataset contains 4,699 malicious code files, including various types of packages, single files, and code snippets. This dataset can be used for subsequent in-depth analysis and research to better understand the properties and behavior of malicious code. Building upon this, we have made the collected dataset publicly available on GitHub, allowing a broader range of researchers to access and utilize the data for further investigation and understanding of malicious code properties and behaviors.

\subsection{Malicious code classification}

\begin{figure}[t!]
  \vspace{-0.5pt}  
  \centering
  \includegraphics[scale=0.25]{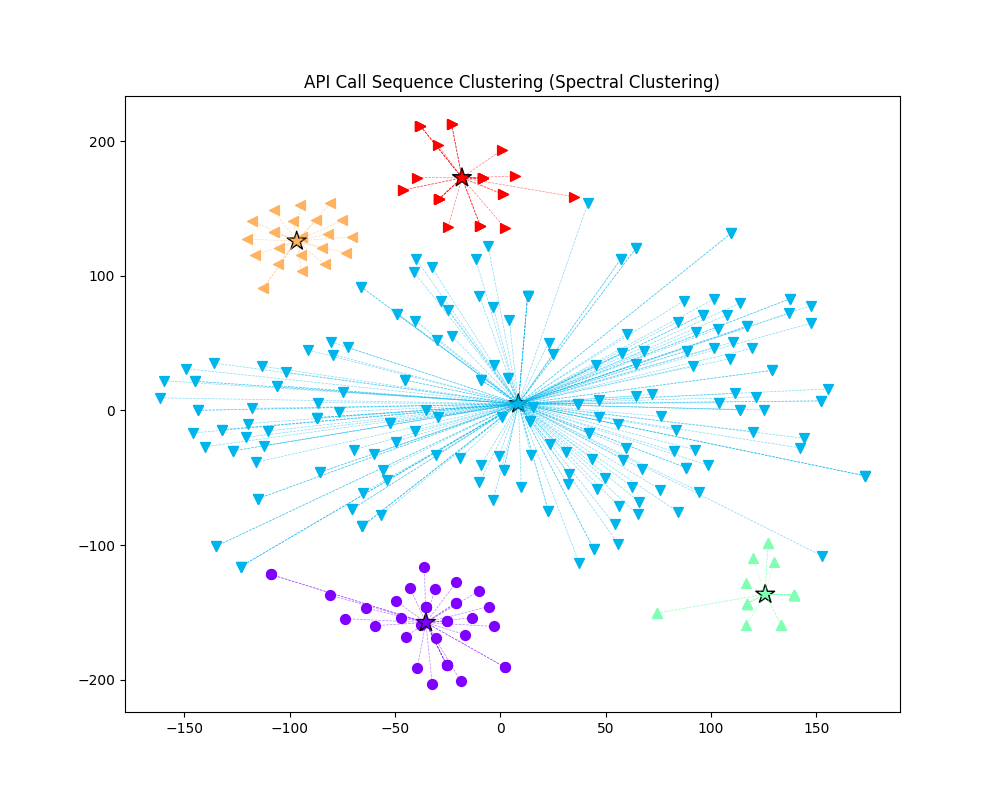}
  \caption{API sequence clustering result}
  \label{clustering}
  \vspace{-1pt}  
\end{figure}

We proposed a clustering method based on API call sequences to analyze and classify malicious code. Our goal is to reveal the behavioural patterns of malware and thus identify malicious code with similar behavioural attributes more effectively. We extract API call sequences from known malicious code closely related to malicious behaviours to achieve this goal. These API call sequences contain essential information about how the malicious code interacts with the operating system or other software modules during execution
By clustering the sequence of API calls, we can discover the common behaviour among different malicious code. Before implementing the clustering approach, we used lexical analysis techniques to tag segment the API calls. This step consists of splitting the code of API calls into basic syntactic elements (e.g., function names, parameters, and operators) that provide input for the subsequent vectorization and clustering process. Next, we use word embedding techniques to convert the sequence of API calls after lexical analysis into numerical vectors. We use the Word2Vec method to convert the original API call sequences into numerical vectors to apply clustering algorithms. Finally, the Spectral Clustering algorithm is used to classify the malicious code into different categories according to their behavioural characteristics. \camera{At the same time, the final classification results are manually confirmed to ensure the accuracy of the clustering results.} As shown in Figure \ref{clustering}, we show the clustering results of API call sequences. We successfully identified five categories with different malicious behaviours. Each category has unique behavioural patterns, which can help us better understand the behavioural characteristics of malware and provide valuable information for further malicious code analysis and detection. \wenboreview{Our website\footnote{\textbf{Detailed Explanation:} \href{https://sites.google.com/view/pypiempircal}{https://sites.google.com/view/pypiempircal}\label{first-footnote}} provides a detailed explanation of five types of malicious behaviour and presents their corresponding formal expression patterns.}

\section{Empirical Study}
\subsection{Life cycle portrait of malicious code}

\begin{figure}[htbp]
  \centering
  \includegraphics[scale=0.55]{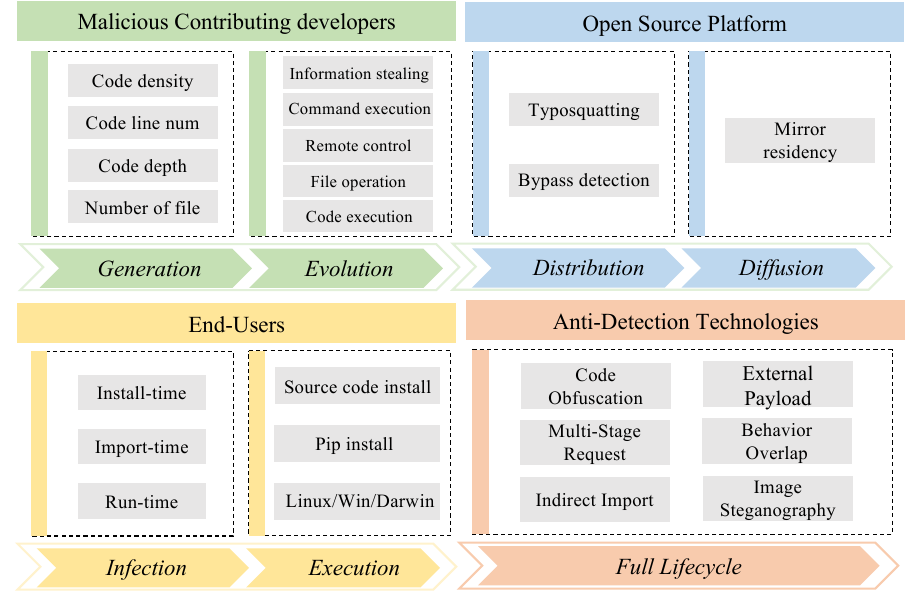}
  \caption{Life cycle portrait of malicious code in PyPI ecosystem}
  \label{portrait}
\end{figure}

In this study, we conducted a large-scale empirical analysis of the constructed multi-source Python malicious code dataset. Finally, we sketched the malicious code lifecycle\cite{shahzad2019large} portrait shown in Figure \ref{portrait}. There are three main parties involved in the lifecycle of the PyPI malicious registry: malicious contributing developers, open source platforms, and end-users. The malicious contributing developers create and continuously evolve the malicious registry with specific behaviours;

the open-source platform PyPI is the major media through which the malware is distributed and diffused to other mirror sources; 
and the end-users introduce malicious packages into the project in a specific way and trigger their malicious behaviours. The malicious code portrait reflects the characteristics of malicious code at different lifecycle stages.

This study aims to: (1) Deeply explore the features and attributes of malicious code in the PyPI ecosystem to reveal its commonalities and differences; (2) Analyze the attack behaviour and evolution characteristics of malicious code to improve the understanding of its behaviour and purpose; (3) Evaluate the detection effectiveness of existing detection tools and study the distribution of malicious code in the wild to stop its propagation more effectively; (4) Study the way malicious code is triggered in end-users to reveal its impact on users; (5) Provide new insights and directions for future research in related fields. In this section, we design appropriate research methods for the four research questions and thoroughly analyze the empirical results. 

\subsection{\textit{RQ1 (Code Attributes): What are the primary attributes and sources of malicious code in the PyPI ecosystem, and how do they compare to other platforms?}}

\textbf{Study Methodology:} To answer RQ1, this study uses the number of files, file depth and malicious code density of Python malicious code as metrics to study the properties of malicious code from different sources and the differences between them. This research focuses on malicious files from the PyPI ecosystem, where malicious code is often hidden between benign code. In order to measure the malicious code density, this study refers to the existing research methods\cite{liu2021characterizing}. It defines LOM/LOC as the malicious code density, where LOM represents the number of lines of malicious code in the project, and LOC represents the total number of lines of Python code in the project. In addition, this study also considers the file depth of the malicious code in the project to analyze its structural characteristics. By analyzing the properties of the malicious code, this study can provide a rough estimate of the size and complexity of the malware and help to understand how the malicious code hides and how to detect it effectively.

\begin{tcolorbox}[colback=gray!5!white, boxrule=0.5pt,colframe=white,size=title,breakable,boxsep=1mm,before={\vskip1mm}, after={\vskip0mm}]
\textbf{Finding 1:} 
\camera{Within the PyPI ecosystem, malicious packages exhibit a relatively lower volume of malicious code; however, they demonstrate a notably prevalence of high code density.}
\end{tcolorbox}

We compared the basic properties of the malicious code in different data sources, including the number of Python files (Py C), the total number of lines (Total L), the average number of lines (Avg L), and the depth of malicious code files in the package (Max D). As shown in Table \ref{property}, in the PyPI data source, the scale of malicious code is relatively small, with an average of 2 Python files per package, and the depth of malicious code files does not exceed 3 layers. In contrast, malicious projects on GitHub exhibit higher complexity, with more malicious lines and larger code sizes, averaging 23 files and a maximum file depth of up to 17. Malicious code in data sources such as Exploit DB and StackOverflow is smaller and less complex. On the other hand, the VX Heavens data source exhibits a higher complexity of malicious code.
The maximum depth of malicious code in PyPI is limited to 3, which is attributed to the package structure. In PyPI, the most effective method of injecting malicious code is by placing it within specific files, thus resulting in a reduced file depth for the malicious code. In contrast, the malicious code on GitHub are often relatively large and complex sample projects. Their functions and scale are relatively large, making their malicious code relatively complex. Malicious code from sources such as Exploit DB and StackOverflow is primarily designed for specific applications or scenarios, so their scale and complexity are comparatively smaller. VX Heavens mainly collects historical malicious code, which typically has a larger scale and encompasses intricate functionality implementations.

We also found that the malicious code density in malicious package files is usually very high, 52\% of malicious package files have a malicious code density of more than 50\%, and 43\% of malicious package files have a malicious code density of more than 90\%. Malicious package files mainly use package name obfuscation to induce victims to download and execute malicious code. Therefore, these well-designed packages contain only malicious code and some standard setup and configuration files required by PyPI. At the same time, we found that the number of malicious code lines in all malicious registries is less than 115 lines. The malicious behaviour of the malicious package is relatively simple, and the complex malicious behaviour is stored in the payload and is downloaded during run-time, not by running the malicious source code directly, which shows that less malicious code is easier to hide in the package files.

\begin{table}
  \caption{Unique properties of malicious code}
  \setlength{\tabcolsep}{1.3pt}
  \footnotesize
  \renewcommand\arraystretch{1.3}
  \label{property}
  \centering
  \begin{tabular}{p{2cm}<{\raggedright}p{1cm}<{\centering}p{1cm}<{\centering}p{1cm}<{\centering}p{1cm}<{\centering}p{1cm}<{\centering}}
  \toprule
  \textbf{Data Source} &\textbf{Pkg C} &\textbf{Py C} &\textbf{Total L} &\textbf{Avg L} &\textbf{Max D} \\
  \midrule
   \textbf{PyPI}   &   2,035   &   3,355   &   1,119,707   &   334   &   3    \\
  
    \textbf{GitHub}   &   2,355   &   54,460   &   10,685,811   &   196   &   17    \\
    
    \textbf{Exploit DB}   &   25   &   25   &   806   &   32   &   1    \\
    
    \textbf{StackOverflow}   &   116   &   116   &   2,283   &   20   &   1    \\
    
    \textbf{VX Heavens}   &   17   &   64   &   14,684   &   229   &   4    \\
    
    \textbf{Attck Tools}   &   1   &   1   &   1   &   1   &   1    \\
    
    \textbf{Hackerforums}   &   1   &   1   &   159   &   159   &   1    \\
  
  \bottomrule
  \end{tabular}
\end{table}

\begin{tcolorbox}[colback=gray!5!white, size=title,breakable,boxsep=1mm,colframe=white,before={\vskip1mm}, after={\vskip0mm}]
\textbf{Finding 2:} While malicious code from different data sources has low similarity and little overlap, we found that attackers use penetration tools to generate malicious code and inject it into the PyPI ecosystem.
\end{tcolorbox}

Detailed similarity analysis showed that malicious code from different data sources did not show prominent propagation characteristics. However, in the PyPI ecosystem, we found some malicious code generated by penetration tools. \camera{Attack tools adopt dedicated templates for generating such malicious code. This specific pattern makes a single sample representative.} Notably, the code within the malicious package "disutil-1.0" is highly similar in behaviour and pattern to the reverse shell code generated by the Metasploit tool\cite{2023metasploit} and also uses \textit{Base64()} encoding is used for obfuscation. This situation shows that malicious code authors will use existing tools to generate attack code and inject them into open-source platforms to expand the spread of malicious code.

\begin{tcolorbox}[colback=gray!5!white, boxrule=0.5pt,colframe=black!75,size=title,breakable,boxsep=1mm,before={\vskip1mm}, after={\vskip0mm}]
\textbf{Answer to RQ1:} \camera{Within the PyPI ecosystem, malicious code exhibits a distinct combination of attributes, characterized by relatively low complexity, yet notably high code density, alongside low similarity with other data sources.}
\end{tcolorbox}

\begin{tcolorbox}[colback=gray!5!white, boxrule=0.5pt,colframe=black!75,size=title,breakable,boxsep=1mm,before={\vskip1mm}, after={\vskip0mm}]
\textbf{Insight:} According to the characteristics of malicious code within the PyPI ecosystem, it is essential to research and develop targeted detection tools. Concurrently, paying attention to attack code generated by penetration tools and the strategies and dissemination methods of malicious code authors can contribute to more effectively identifying and mitigating potential security threats.
\end{tcolorbox}

\subsection{\textit{RQ2 (Attack Tactics): How do attackers combine various attack strategies and malicious behaviors when injecting code into open-source packages in the PyPI ecosystem, and how do these tactics evolve and adapt across different platforms and objectives?}}

\textbf{Study Methodology:} To understand the true intentions of malicious contributing developers, we extract the API call sequences of malicious code and cluster their malicious behaviours. The same malicious behaviours have similar API patterns. At the same time, to discover the detailed design of the malicious code, we use a semi-automatic method to audit the code. First, we manually audit some malicious code and extract the behavioral pattern. Then, we use the similarity matching method to filter the remaining malicious packages to discover the malicious code and the files that have been injected, \camera{and we manually check the classification results to ensure the accuracy of the classification results.}

\begin{tcolorbox}[colback=gray!5!white, size=title,breakable,boxsep=1mm,colframe=white,before={\vskip1mm}, after={\vskip0mm}]
\textbf{Finding 3:} Within the PyPI ecosystem, multiple malicious behaviours are prevalent in malicious code, with information stealing and command execution behaviours especially prominent, which indicates that attackers prefer comprehensive methods to achieve their attack objectives.
\end{tcolorbox}

In an in-depth study of 1,335 Python packages, we found that the malicious code exhibited multiple malicious behavioural characteristics, as shown in Table \ref{malicious_behaviour}. Specifically, 630 packages exhibited only one malicious behaviour, while another 635 malicious packages simultaneously had two or more malicious behaviours. Among these malicious packages, 411 packages exhibited two malicious behaviours. We observed that information stealing and command execution behaviours were particularly prevalent. Detailed data revealed that 49.44\% of the packages involved information stealing. However, among them, only 4 packages were related to browser information stealing, with the rest targeting system information stealing. Additionally, 59.33\% of the packages contained command execution behaviour. In addition, 2.62\% of the packages involved code execution, 0.75\% involved remote control, and 47.19\% involved unauthorized file operations. Notably, 7.87\% of the packages used obfuscation techniques to conceal the malicious code to increase the attack's stealthiness. This finding indicates that attackers continuously improve their strategies to evade security defences and detection.

\begin{table}
  \caption{Malicious Behaviors in Packages}
  \setlength{\tabcolsep}{1.3pt}
  \footnotesize
  \renewcommand\arraystretch{1.3}
  \label{malicious_behaviour}
  \centering
  \begin{tabular}{p{2.7cm}<{\raggedright}p{1.3cm}<{\centering}p{2cm}<{\centering}p{1.3cm}<{\centering}}
  \toprule
  \textbf{Behaviors} &\textbf{Package quantity} &\textbf{Behaviors} &\textbf{Package quantity} \\
  \midrule
    \textbf{Command Execution}   &   792   &   \textbf{Code Execution}   &   35   \\
    \textbf{Information Stealing}   &   664   &   \textbf{Remote Control}   &   10  \\
    \textbf{File Operation}   &   630   &      &    \\
  \bottomrule
  \end{tabular}
\end{table}

Attackers tend to use various methods to achieve their attack goals. Unauthorized file operations are often used to implement more complex malicious behaviours, such as reading locally sensitive files and installing Trojan programs. Command execution mainly implements behaviours such as information stealing and permission modification by executing malicious commands. System information stealing is common because system information (such as system version, patch programs, Etc.) can provide attackers with valuable information and facilitate subsequent attacks. The relatively low proportion of browser information stealing may be because malicious behaviours are more targeted and limited by factors such as the browser type and version users use.

\begin{tcolorbox}[colback=gray!5!white, size=title,breakable,boxsep=1mm,colframe=white,before={\vskip1mm}, after={\vskip0mm}]
\textbf{Finding 4:} Malicious code show significant pertinence and adaptability, employing distinct attack strategies and trigger mechanisms tailored to different platforms.
\end{tcolorbox}

To accurately trigger malicious code on various operating systems, attackers usually first use the "\textit{sys.platform}" to obtain the victim's system information, then carefully design adaptable malicious code for different systems. In our research, we found the number of malicious packages targeting win32 was as high as 428, far more than win64 (25), Linux (43), and Darwin (19).
Moreover, we observed that malicious code exhibits a high degree of stealthiness. After downloading and installing trojan files, some malicious code will actively erase installation traces to evade detection. Attackers often utilise methods such as "\textit{os.remove}" and "\textit{shutil.rmtree}" to achieve this goal, as well as using command lines like "\textit{cmd /c del}" to delete trojan files. This heightened concealment indicates that attackers are becoming increasingly sophisticated in their tactics, intending to make malicious code more challenging to detect and analyse.

\begin{tcolorbox}[colback=gray!5!white, size=title,breakable,boxsep=1mm,colframe=white,before={\vskip1mm}, after={\vskip0mm}]
\textbf{Finding 5:} Malicious code in the PyPI ecosystem is not static but constantly evolving. Attackers continue to improve their malicious behaviour to achieve more complex and refined functionality and code structures.
\end{tcolorbox}

In the PyPI ecosystem, malicious code does not always remain static. Attackers continue to optimize and update these code to achieve more complex and diverse malicious behaviours. Among the collected malicious packages, we found that the malicious code in 35 versions of 18 packages had evolved behaviours. These evolutionary trends are mainly manifested in the diversification of malicious behaviours and the simplification and efficiency of code structures.
Taking "dpp\_client" as an example, in later versions, the attacker incorporated exception-handling mechanisms to enhance the robustness of the malicious code, enabling it to better cope with issues when encountered. Another example is the "Collored" malicious package, whose primary function is downloading trojan files from a remote server and installing and executing them locally. From the initial version 0.0.3 (comprising 26 lines of code) to version 0.0.5 (containing 24 lines of code), and finally to version 0.0.7 (including just 17 lines of code), the malicious code was progressively condensed, making the download and installation process of the trojan files more efficient. However, despite the changes in the code structure, the core functionality remained unchanged. We also discovered that "easyfuncsys" could evade the detection of security tools during the evolution of its subsequent versions.

\begin{tcolorbox}[colback=gray!5!white, boxrule=0.5pt,colframe=black!75,size=title,breakable,boxsep=1mm,before={\vskip1mm}, after={\vskip0mm}]
\textbf{Answer to RQ2:} \camera{Attackers adeptly integrate diverse attack strategies and malicious behaviors, with information stealing particularly pronounced among these multifaceted malicious behaviors. Malicious code exhibits distinct pertinence and adaptability, featuring varying adaptation codes for different systems. Moreover, malicious code demonstrates a discernible trait of continuous evolution, aimed at achieving heightened complexity and refined functional.}
\end{tcolorbox}

\begin{tcolorbox}[colback=gray!5!white, boxrule=0.5pt,colframe=black!75,size=title,breakable,boxsep=1mm,before={\vskip1mm}, after={\vskip0mm}]
\textbf{Insight:} Deepen the understanding of the underlying reasons for malicious behaviour and the purposes of attackers, and focus on the evolutionary characteristics of malicious code to develop efficient detection tools.
\end{tcolorbox}

\subsection{\textit{RQ3 (Evasion Techniques and Distribution): How effective are existing detection tools at identifying malicious packages, which evasion methods do malicious code employ to elude these tools, and what is the impact of these methods on their distribution in real-world applications?
}}

\textbf{Study Methodology:} This study focuses on an in-depth analysis of how malicious code utilises various strategies and techniques to evade detection by security detection tools. To achieve a comprehensive and unbiased comparison, we selected multiple representative malicious code detection tools in Table \ref{tools_description}, which are based on different principles and techniques to identify malicious code in projects. We constructed the evaluation dataset shown in Table \ref{evalution_dataset}, which includes 1,556 malicious samples from PyPI and 549 randomly selected benign samples, including some popular software packages. These samples cover different malicious behaviours and techniques, such as information stealing, Trojan download and installation, remote access, etc. We use these samples to evaluate the performance and detection capabilities of different malicious code detection tools.

\begin{table}
  \caption{Benchmark dataset for Analysis}
  \setlength{\tabcolsep}{1.3pt}
  \footnotesize
  \renewcommand\arraystretch{1.3}
  \label{evalution_dataset}
  \centering
  \begin{tabular}{p{1.5cm}<{\raggedright}p{1.5cm}<{\centering}p{2cm}<{\centering}p{2cm}<{\centering}}
  \toprule
  \textbf{Dataset} &\textbf{Packages} &\textbf{Python Files} &\textbf{Lines of Code} \\
  \midrule
    \textbf{Malicious}   &   1,556   &   6,887   &   254,242    \\
    \textbf{Benign}   &   549   &   7,762   &   618,020     \\
  \bottomrule
  \end{tabular}
\end{table}

\begin{table}
  \caption{Python malicious code detection tools}
  \setlength{\tabcolsep}{1.3pt}
  \footnotesize
  \renewcommand\arraystretch{1.3}
  \label{tools_description}
  \centering
  \begin{tabular}{p{2cm}<{\raggedright}p{3.5cm}<{\centering}p{2.5cm}<{\centering}}
  \toprule
  \textbf{Tool name} &\textbf{Detection Principle} &\textbf{Available }\\
  \midrule
    \textbf{Bandit4Mal}   &   pattern matching, rule-based   &   source code   \\
    \textbf{OSSGadget}   &   regex rule   &   package, source code    \\
    \textbf{Aura}   &   static analysis, rule-based   &   binary, source code  \\
    \textbf{PyPI Check}   &   yara rule   &   package, source code     \\
    \textbf{VirusTotal}   &  feature, dynamic analysis  & package, source code    \\
    \textbf{Pyt}   &   static analysis   &   source code     \\
    \textbf{Snyk Code Test}   &  AI-based semantic Analysis  &  package, source code  \\
    \textbf{ClamAV}   &  signature database  &  source code  \\
  \bottomrule
  \end{tabular}
\end{table}

\begin{tcolorbox}[colback=gray!5!white, size=title,breakable,boxsep=1mm,colframe=white,before={\vskip1mm}, after={\vskip0mm}]
\textbf{Finding 6:} Existing malicious code detection tools demonstrate some effectiveness in identifying malicious packages, but the false-positive rate remains considerably high.
\end{tcolorbox}

\begin{figure}[t]
  \vspace{-0.5pt}  
  \flushleft
  \includegraphics[scale=0.6]{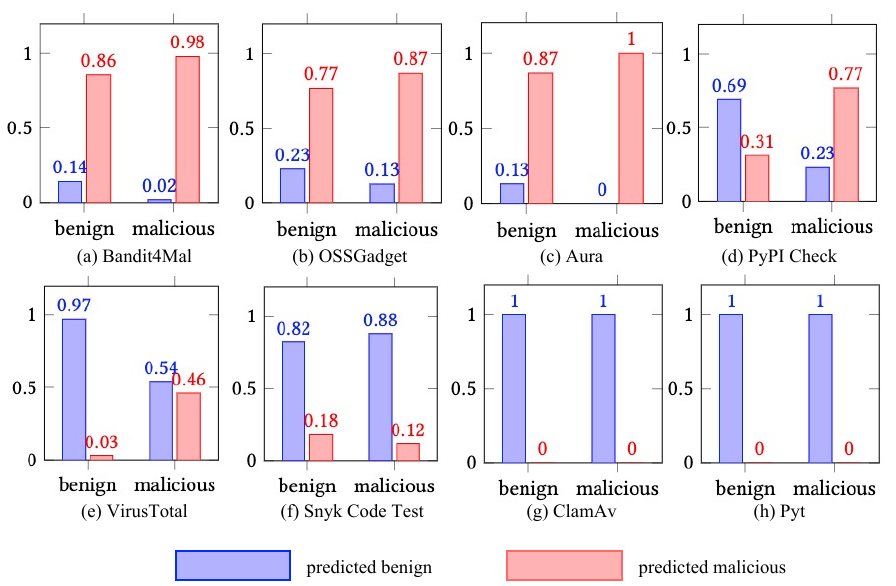}
  \caption{Detection results of different tools}
  \label{tool_result}
  \vspace{-0.5pt}  
\end{figure}

Figure \ref{tool_result} illustrates the detection performance of various tools, indicating that current detection tools for malicious Python packages suffer from severe issues in terms of accuracy. We found that Bandit4mal has a false positive rate as high as 86\% when detecting benign samples, with only 14\% of benign samples being correctly identified; OSSGadget-backdoor has a false positive rate of 77\%, while the proportion of accurately detected benign samples is merely 23\%. Similarly, the performance of these tools in detecting malicious packages is also disappointing. For instance, Aura has a false positive rate of up to 87\%, whereas Virustotal has an astonishing false negative rate of 54\%. Comparatively, the PyPI Check performs slightly better, but due to its detection being limited to the setup.py file, its false negative rate reaches 23\%. Additionally, we discovered significant disparities among different security detection tools in terms of false negatives. For example, OSSGadget and PyPI Check failed to detect numerous malicious code related to information stealing and unauthorized file operations. At the same time, VirusTotal missed many malicious packages associated with command execution.

After an in-depth analysis of the detection principles of these tools, we believe that the reasons for this phenomenon are mainly attributed to the following points: First, rule-based and pattern-based detection methods (such as Bandit4mal, OSSGadget and Aura) tend to misclassify behaviours such as network connections and file operations in normal packages as malicious behaviours during the detection process. The root cause of this misjudgment is that these tools have difficulty distinguishing the subtle difference between malicious behaviour and normal behaviour. Second, signature-based detection methods (such as Virustotal and ClamAv) perform poorly in dealing with diverse and constantly changing malicious code because the fingerprints of malicious code are challenging to maintain consistency, making it difficult for signature databases to capture the latest malicious packages. Finally, some tools (such as PyPI Check) have limitations in the scope of detection and only detect ``setup.py'', which leads to false negatives when dealing with malicious code of import-time and run-time attacks. In response to the fact that OSSGadget and PyPI Check mainly miss malicious code related to information stealing and unauthorized file operation, \wenboreview{we believe these tools mainly focus on identifying backdoors and other obvious malicious behaviours, making it difficult to detect more covert malicious activities.} In addition, these tools are rule-based detection methods. These tools cannot deeply analyze the execution path of the code and the interaction with external resources, making it difficult to detect malicious behaviour. The signature database of VirusTotal cannot contain all variant forms of malicious code so that the malicious code can evade detection by general obfuscation or other anti-detection techniques.

\begin{tcolorbox}[colback=gray!5!white, size=title,breakable,boxsep=1mm,colframe=white,before={\vskip1mm}, after={\vskip0mm}]
\textbf{Finding 7:} Malicious code employs numerous anti-detection techniques to evade security tools, including code obfuscation, external payloads, multi-stage requests, behaviour overlap, indirect import, image steganography, and sandbox escape.
\end{tcolorbox}

For conventional code obfuscation techniques, the hidden logic and intent in malicious code can be found by analyzing the API in the abstract syntax tree. However, some malicious code use advanced obfuscation techniques, such as anti-debugging and anti-virtualization, which are often difficult to discover through program analysis tools. In this case, we need to use manual analysis to check the malicious code and explore the anti-detection technology carefully. While analyzing the malicious code, we found that attackers use multiple methods to evade the detection of security tools. \wenboreview{We summarize seven anti-detection techniques, including code obfuscation, external payload, multi-stage request, behaviour overlap, indirect import, image steganography, and sandbox escape. Among them, indirect import, image steganography and sandbox escape are techniques we discovered for the first time in Python malicious code studies. This paper will focus on these three techniques, while detailed information about the other techniques is available on our website.\footref{first-footnote}.}

\textbf{Indirect Import}: In order to bypass the detection tool, the attacker does not directly inject malicious code into the current package but implements the attack by indirect reference, as shown in Listing \ref{code5}. ``secbg-0.0.8'' introduces the malicious package ``secrevtwo-0.0.1'' via ``\textit{install\_requires}'' in the ``setup.py'' file.

\begin{lstlisting}[style=customPython, caption={Indirect Import in secrevtwo-0.0.3.}, label=code5]
setup(install_requires=['secrevtwo']
try:
  p = subprocess.Popen(["python3", "-c", \
  "from secrevtwo import dist_util"],close_fds=True)
\end{lstlisting}

\textbf{Image Steganography}: Attackers use steganography to hide malicious code in an image, and when the package code is executed, it downloads the image from a remote server and extracts the malicious code from it for execution. ``colorsapi-6.6.7'' (Listing \ref{code6})exploits this type of attack.

\begin{lstlisting}[style=customPython, caption={Image Steganography in colorsapi-6.6.7.}, label=code6]
with open(f'{os.getenv("TEMP")}\\a.png','wb') as f:
    f.write(r.content)
exec(lsb.reveal(f'{os.getenv("TEMP")}\\a.png'))
\end{lstlisting}

\textbf{Python Sandbox Escape}: attackers usually use the \textit{\_\_import\_\_} method to dynamically import the required module and compile and execute malicious code at runtime (Listing \ref{code7}) with the help of built-in functions \textit{builtins}. According to the data we collected, 63 packages related to sandbox escape have been identified.

\begin{lstlisting}[style=customPython, caption={Python Sandbox Escape in colorwin-6.6.7.}, label=code7]
__import__('builtins').exec(__import__('builtins').
compile(__import__('base64').b64decode("..."),
'<string>','exec'))
\end{lstlisting}

Image steganography technology can make malicious code challenging to be detected and analysed, while indirect import technology can make malicious code difficult to be found. These findings further prove that attackers use increasingly sophisticated and varied techniques to evade detection by security detection tools.

\begin{tcolorbox}[colback=gray!5!white, size=title,breakable,boxsep=1mm,colframe=white,before={\vskip1mm}, after={\vskip0mm}]
\textbf{Finding 8:} Many malicious packages exist in PyPI mirror servers in various countries, among which China's mirror ecosystem is the most severe.
\end{tcolorbox}

\begin{table}
  \caption{Distribution of malicious packages in the mirrors}
  \setlength{\tabcolsep}{1.3pt}
  \footnotesize
  \renewcommand\arraystretch{1.3}
  \label{ablation}
  \centering
  \begin{tabular}{p{1.5cm}<{\raggedright}p{3cm}<{\raggedright}p{2.3cm}<{\raggedright}p{1.5cm}<{\raggedright}}
  \toprule
  \textbf{Country} & \textbf{Registry Mirrors} &\textbf{Package Nums} &\textbf{Percentage} \\
  \hline
  \multirow{1}{*}{} 
	& \textbf{Total} &  2,035 &  100\%  \\
	\hline
  \multirow{7}{*}{\textbf{CN}} 
   & \textbf{Tsinghua U} \cite{tsinghua}   &   1,661  &   81.62\%   \\
   & \textbf{Douban} \cite{douban}   &   369   &  18.13\%   \\
   & \textbf{Huawei} \cite{huawei}   &   88   &  4.32\%   \\
   & \textbf{BFSU} \cite{bfsu}   &   1,638   &  80.49\%   \\
   & \textbf{NetEase} \cite{netease}   &   53   &  2.60\%   \\
   & \textbf{Aliyun} \cite{alibaba}   &   5   &  0.25\%   \\
   & \textbf{Tencent} \cite{tencent}   &   1,362   &  66.93\%   \\
   & \textbf{Sustech} \cite{sustech}   &   450   &  22.11\%   \\
    \hline

 \multirow{3}{*}{\textbf{US}}
   & \textbf{Rstudio} \cite{rstudio}   &   31   &  1.52\%   \\
   & \textbf{3.225.43.100}    &   2   &  0.10\%   \\
   & \textbf{3.217.35.114}    &   4   &  0.20\%   \\
    \hline
 \multirow{1}{*}{\textbf{HK}}
   & \textbf{101.33.123.191}    &   71   &  3.49\%   \\
    \hline
 \multirow{1}{*}{\textbf{KO}}
   & \textbf{Kakao} \cite{kakao}   &   224   &  11.01\%   \\
    \hline
 \multirow{1}{*}{\textbf{ID}}
   & \textbf{Padjadjaran U} \cite{unpad}   &   7   &  0.34\%   \\
\bottomrule
  \end{tabular}
\end{table}

In the previous section, we evaluated the performance of detection tools and found that many tools cannot effectively detect malicious packages in PyPI. Based on this, we speculate that the number of malicious packages in actual application scenarios may be much larger than the known data. To test this hypothesis, we conducted a study of the distribution of malicious packages in the PyPI ecosystem in the wild. The results show that there are a large number of malicious packages in the PyPI mirror sites around the world. Table \ref{ablation} shows the distribution of malicious packages in PyPI mirror servers in various countries. This is especially evident in mirror servers in China (CN), covering different PyPI mirrors such as Tsinghua, Douban, BFSU, Netease, Alibaba Cloud, and Tencent. In addition, mirror servers in the United States (US), such as Rstudio and two IP addresses (3.225.43.100 and 3.217.35.114), also have a certain amount of malicious packages. We also found malicious packages in mirror servers in Hong Kong (HK), South Korea (KO) and Indonesia (ID), such as 101.33.123.191 in Hong Kong, Kakao in South Korea, and Padjadjaran University in Indonesia (Padjadjaran U).
In addition, we discovered that 70 versions of 37 malicious packages were not present in the 14 mirrors. However, in the file\_downloads records, the download history of these malicious packages still exists. They have been downloaded within the past six months, which indicates that even if these malicious packages do not appear in these 14 mirror sites, there are still many other mirror sites within the PyPI ecosystem where they may continue to pose a threat to end users. Due to the sheer number of these sites, it is difficult for us to track the specific distribution of these malicious packages.

The widespread distribution of malicious code in PyPI mirrors relates to the way mirrors are constructed. PyPI mirrors are built by periodically synchronizing packages from the root PyPI repository to mirror servers. Consequently, if a malicious package is uploaded to the root PyPI repository, it will be synchronized to the mirrors, leading to the propagation of malicious code. Simultaneously, the synchronization frequency between the root repository and mirrors may differ, resulting in outdated packages in some mirrors. 
Moreover, even if the root site removes a malicious package, some mirror construction methods (such as Bandersnatch) may cause local mirrors not to synchronize the deletion, allowing malicious packages to persist on mirror sites for extended periods.

\begin{tcolorbox}[colback=gray!5!white, boxrule=0.5pt,colframe=black!75,size=title,breakable,boxsep=1mm,before={\vskip1mm}, after={\vskip0mm}]
\textbf{Answer to RQ3:} \camera{Malicious code has proven adept at evading detection from existing tools through the employment of numerous anti-analysis techniques. Concurrently, a substantial presence of malicious packages persists within the PyPI repository, posing a severe security threat to end-users.}
\end{tcolorbox}

\begin{tcolorbox}[colback=gray!5!white, boxrule=0.5pt,colframe=black!75,size=title,breakable,boxsep=1mm,before={\vskip1mm}, after={\vskip0mm}]
\textbf{Insight:} It is necessary to design targeted de-obfuscation and decryption methods to counter the anti-detection techniques used by malicious code. Additionally, mirror maintainers should promptly remove malicious packages to enhance the security of the mirror.
\end{tcolorbox}

\subsection{\textit{RQ4 (Impact and Infiltration) Within the PyPI ecosystem, how has the impact of malicious packages on end-users evolved over time, which operating systems have been affected, and what methods have been employed by malicious packages to infiltrate user systems?}}

\textbf{Study Methodology:} The lifecycle of a vulnerability is usually described as the stages of vulnerability discovery, utilization, propagation, patching, and prevention. Similarly, we define the lifecycle of a malicious registry as the time from when a developer uploads it to PyPI to when it is completely deleted. 
Due to out-of-sync mirror sources, the lifecycle of these malware packages may be extended. 
To analyze the harm caused by malicious registry files to users, we collected the download records (file\_downloads) and metadata (meta-data) of 861 PyPI registry files from Google Cloud. We extracted the release time, first download time and the last download time. Some data are problematic, so we compare the release time with the first download time and use the smaller value of the two as the release time. The discovery time of the malicious registry comes from the open-source database. We define the time between malicious registry release and malicious registry discovery as the incubated period, and the time between malicious registry discovery and the last download of the registry as the residual period.

\begin{tcolorbox}[colback=gray!5!white, size=title,breakable,boxsep=1mm,colframe=white,before={\vskip1mm}, after={\vskip0mm}]
\textbf{Finding 9:} Many malicious registries (72.34\%) have a long residual period beyond their incubated period after discovery, which highlights that these malicious registries still persist after being discovered.
\end{tcolorbox}

\begin{figure}[htbp]
  \centering
  \includegraphics[scale=0.13]{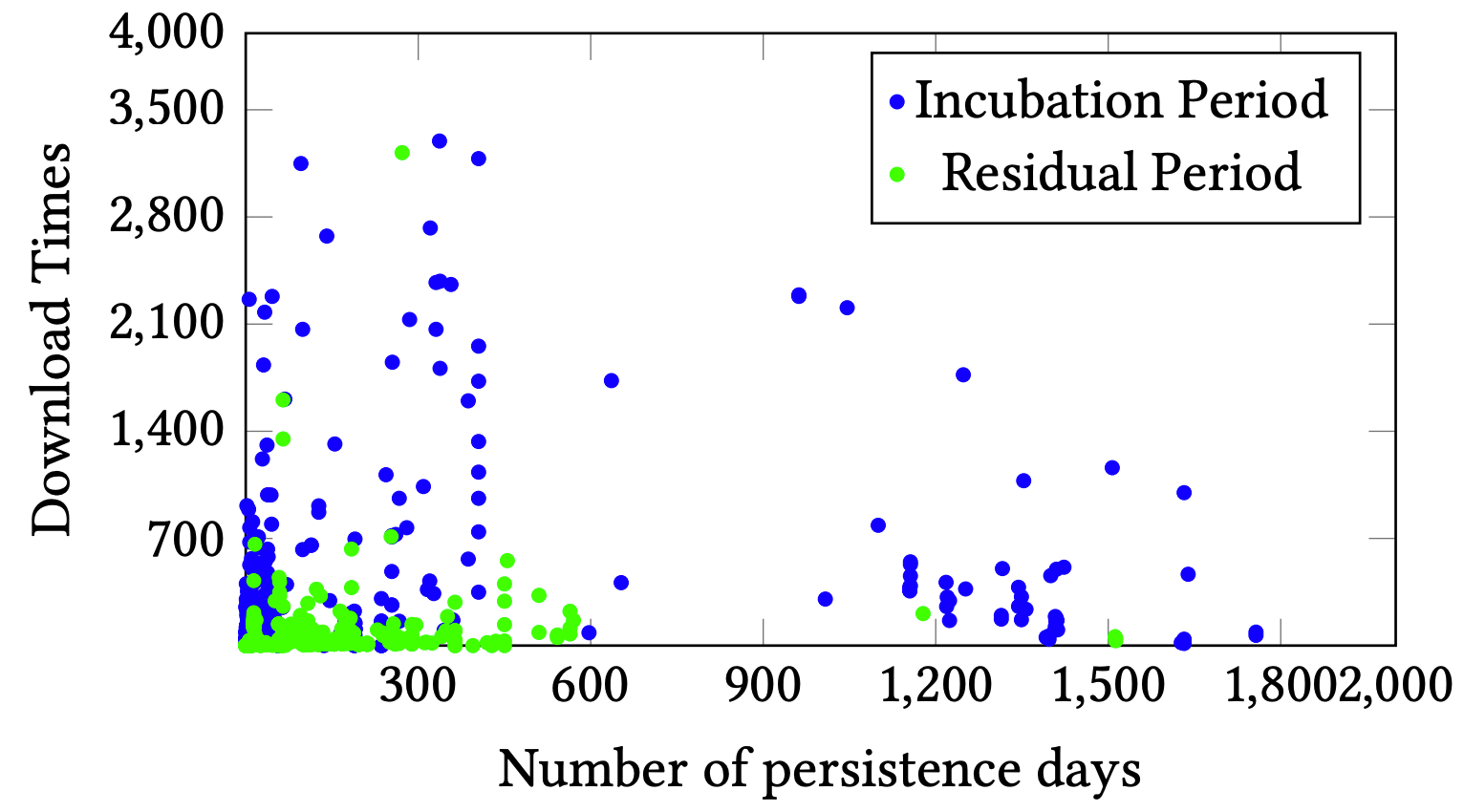}
  \caption{Scatter figure of malicious package download times during the incubation period and residual period}
  \label{scatter}
\end{figure}

Our findings reveal that 622 malicious packages had a residual period longer than the incubation period, accounting for 72.34\% of the total. Also we found that 28\% of the packages had a incubation period of more than 2 months, and this figure rose to 40\% in the residual period, 39\% and 31\% of the packages were downloaded more than 100 times during the incubation and residual periods, respectively, and 6\% of the malicious packages were downloaded more than 1,000 times. Notably, the ``pyscrapy'' malicious package had an incubation period of 5 years, while the "request'' package was downloaded 6,061,233 times during the incubation period. Furthermore, as shown in Figure \ref{scatter}, ``libpeshka" and ``djanga'' packages had a residual period of 1,500 days, while ``colorwed'' was downloaded 1,605 times during the residual period.

\begin{tcolorbox}[colback=gray!5!white, size=title,breakable,boxsep=1mm,colframe=white,before={\vskip1mm}, after={\vskip0mm}]
\textbf{Finding 10:} There are three attack vectors in the PyPI ecosystem, Install-Time attacks, Run-Time attacks, and Import-Time attacks, among which the most attacks are at Install-Time.
\end{tcolorbox}

We analyzed the collected dataset and found three attack vectors, which are Install-Time Attack\cite{wyss2022wolf}, Import-Time Attack and Run-Time Attack.

\begin{figure}[htbp]
  \centering
  \includegraphics[scale=0.65]{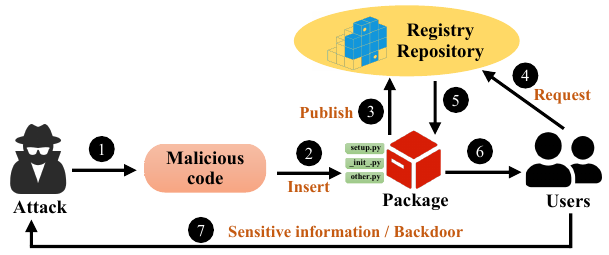}
  \caption{Overview of Package Install-Time Attack}
  \label{install_attack}
\end{figure}

Figure \ref{install_attack} shows the flow of the attack during installation. The attacker inserts the written malicious code into the setup.py file in the package. It is published to the registry repository and is synced to other registry mirrors. When users request a source package file, the attacker-uploaded package embedded with malicious code will be downloaded. \wenboreview{Specifically, the attacker reimplemented an installation class \textit{CustomInstall}, and then rewrite the \textit{run} method in it, and then let \textit{setuptools} call this installation class during installation.} The user executes the installation program locally to trigger the attack, sending sensitive information of the host to the attacker or establishing a backdoor to achieve persistent control.

\begin{figure}[htbp]
  \centering
  \includegraphics[scale=0.6]{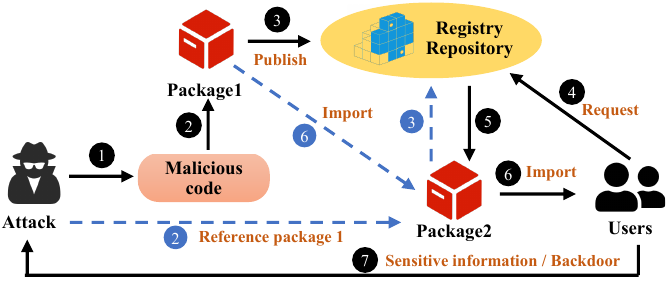}
  \caption{Overview of Package Import-Time Attack}
  \label{import_attack}
\end{figure}

Figure \ref{import_attack} is another way of import-time attack. The attacker first inserts malicious code into package 1 and uploads it to the registry repository. Then create package 2 and import package 1 in it. When the user imports package 2, package 1 will be downloaded to the project, and the attack will be triggered. This method can evade security detection tools.

Run-Time attacks involve embedding malicious code into files within ``other.py'', allowing for multiple types of attacks to be carried out. The attack is executed by calling the functions containing the embedded malicious code, which will activate when the user runs the code. This type of attack usually occurs in other files except the ``\_\_init\_\_.py'' files within the package. \wenboreview{The pytagora-1.2 malicious package uses this type of attack.}

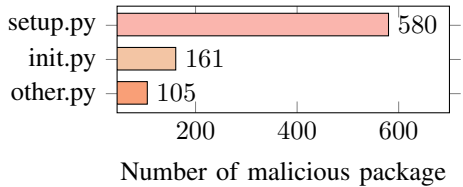
\begin{figure}[htbp]
  \centering
  \definecolor{mycolor1}{RGB}{249,159,119}
  \definecolor{mycolor2}{RGB}{243,197,164}
  \definecolor{mycolor3}{RGB}{250,181,173}
  \begin{tikzpicture}
    \begin{axis}[
        xbar,
        width=6cm,
        height=3cm,
        xlabel={Number of malicious package},
        xmax=700,
        symbolic y coords={other.py,init.py,setup.py},
        bar width=0.3cm,
        enlarge y limits=0.28,
        bar shift=0.01cm, 
        nodes near coords,
        nodes near coords align={horizontal},
    ] 
        \addplot [fill=mycolor1] coordinates {(105,other.py)};
        \addplot [fill=mycolor2] coordinates {(161,init.py)};
        \addplot [fill=mycolor3] coordinates {(580,setup.py)};
    \end{axis}
\end{tikzpicture}
\caption{Distribution of malicious code in the package}\label{code_distribution}
\vspace{-1pt}  
\end{figure}

We analyzed all the malicious PyPI packages collected and the results are shown in Figure \ref{code_distribution}. We found that 580 (68.6\%) of the 846 malicious packages belonged to Install-Time Attack, 161 (19\%) to Import-Time Attack, and 105 (12.4\%) to Run-Time Attack.

\begin{tcolorbox}[colback=gray!5!white, size=title,breakable,boxsep=1mm,colframe=white,before={\vskip1mm}, after={\vskip0mm}]
\textbf{Finding 11:} Malicious packages are mainly installed through source code (74.81\%), with Linux systems being the most affected (77.08\%). Therefore, Linux users need to choose and install packages carefully when using tools such as \textit{pip}, and also verify the source and trustworthiness of the packages carefully.
\end{tcolorbox}

\begin{figure}[htbp]
  \vspace{-1pt}  
  \centering
  \includegraphics[scale=0.5]{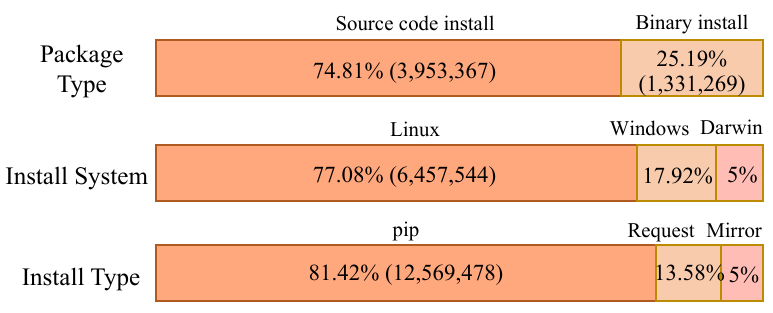}
  \caption{Download statistics analysis of malicious packages}
  \label{impact_system}
  \vspace{-1pt}  
\end{figure}

The results of our study are shown in Figure \ref{impact_system}, malicious packages were most commonly installed via source code, accounting for 74.81\% of the total downloads, or 3,953,367 times. In contrast, binary installation only accounted for 25.19\%, mainly because most malicious packages only provided source code distribution packages. Attackers must install these packages through the source code to trigger the attack during installation. Moreover, our research showed that Linux systems suffered the most serious attacks, with 6,457,544 malicious packages downloaded, accounting for 77.08\% of the total. It highlights the importance of Linux system users being more cautious when selecting and installing packages to avoid being victimized by malware.
Regarding attack methods, we found that 81.42\% of malicious packages were imported to users' systems through \textit{pip}. Therefore, it is essential for users to carefully check the source and trustworthiness of packages when installing them. Additionally, 13.58\% of the malicious packages were installed by \textit{requests}, and 5\% were synchronized to different registry mirrors by mirroring, among which the \textit{Bandersnatch} synchronization mechanism is the primary method. It indicates many malicious packages in mirrors, and users should choose the mirror source carefully.

\begin{tcolorbox}[colback=gray!5!white, boxrule=0.5pt,colframe=black!75,size=title,breakable,boxsep=1mm,before={\vskip1mm}, after={\vskip0mm}]
\textbf{Answer to RQ4:} \camera{In the PyPI ecosystem, the impact of malicious packages on end users has exhibited a gradual increase over time. Malicious packages primarily infiltrate user systems via source code installation attacks, with the Linux system being most affected.}
\end{tcolorbox}

\begin{tcolorbox}[colback=gray!5!white, boxrule=0.5pt,colframe=black!75,size=title,breakable,boxsep=1mm,before={\vskip1mm}, after={\vskip0mm}]
\textbf{Insight:} Researchers should strengthen the monitoring and analysis of malicious packages and improve detection efficiency and accuracy. End-Users should exercise caution when selecting and installing packages using pip and other tools and verify the software packages' sources and credibility to ensure system security.
\end{tcolorbox}

\section{Implications}

Based on our findings, We propose mitigation measures for each type of participant involved and research directions for future works.

\wenboreview{\textbf{Users} need to raise their security awareness. Installing security packages, conducting regular scans of the host system, and promptly updating local software packages are essential. Utilizing official registries and secure mirror sources ensures the reliability and safety of the downloaded software packages. \textbf{Registry Maintainers} should ensure the security of registries and prevent the spread of malicious software packages. This includes banning accounts of malicious maintainers and strengthening review mechanisms. Providing advanced and comprehensive detection tools to identify malicious packages is essential for filtering and defence. Supplying a complete blacklist of malicious packages guarantees the safety of mirror sources. \textbf{Registry Mirrors Maintainers} need to synchronize regularly with the official registry to ensure consistency and avoid missing important updates and fixes. Promptly removing malicious packages ensures the safety of PyPI mirrors. \textbf{Researchers} can delve into attack vectors to explore new attack patterns and techniques, study advanced and comprehensive malicious code detection methods to counteract evasion techniques, focus on code evolution to gain insights into evolution trends and study multi-behavioural features to optimize malicious code identification and prevention. Utilize dynamic and static analysis and machine learning techniques to discover new malicious code.  Focus on install-time and import-time attack vectors to improve scanning efficiency when scanning open-source packages on a large scale.}

\section{Related Work}

\subsection{Malicious Code Detection}

\wenboreview{GitHub and other repositories can be easily used to distribute malicious code and software libraries.} Anomalicious utilizes commit logs and repository metadata to automatically detect anomalies and potentially malicious commits, and successfully detected 8 out of 15 known malicious commits. Attackers use the fork function of GitHub repositories as a storage and distribution channel for malware\cite{gonzalez2021anomalicious}. To address this problem, Cao et al.\cite{cao2022fork} used automated detection and reverse engineering techniques to analyze the artifacts in a given repository. To improve detection efficiency, they built a similarity database to identify potential malware faster. They successfully identified 26 fork repositories containing malware among 68 popular cryptocurrency repositories. Zhang et al.\cite{zhang2020cyber} used the enhanced deep neural network DNN to analyze the code content of the github repository, and used the heterogeneous information network HIN to model the neighborhood relationship to improve the recognition accuracy. Attackers often embed malicious shell commands into Python scripts for illegal operations. However, traditional static analysis methods cannot detect such attacks. Zhou et al. \cite{zhou2022pycomm} proposed a machine learning-based model named PyComm for detecting malicious commands in Python scripts with multidimensional features, which simultaneously considered 12 statistical features of Python source code and string sequences. Fang et al.\cite{fang2021pbdt} used machine learning methods to detect Python backdoors, represented the text through the statistical features caused by confusion and the characteristics of the opcode sequence in the compilation process, and matched suspicious modules and functions in the code. This method can effectively detect embedded backdoor in the code.

\subsection{Package Management Security}

To detect various malicious software packages in the open source software registry. Duan et al.\cite{duan2020towards} constructed a multi-dimensional analysis framework. Analyze the security of the registry from the basic information of the registry, function calls at the source code level, package execution and system calls in dynamic analysis. And detected 339 new malicious packages in PyPI, npm, RubyGems open source software package. 
Gu et al.\cite{gu2022investigating} built a continuous monitoring and analysis framework named RScouter for mainstream package managers, and found 12 potential attack vectors in the continuous monitoring of 6 registries, which can be used by attackers to distribute malicious package files. 
Meanwhile, a large number of suspicious packages were found. Liang et al.\cite{liang2021malicious} proposed a third-party malware library identification framework named PPD based on anomaly detection. First import the packages required by the library to form a complete code package, then use AST and RegExp to extract code features (IP address/dangerous function, etc.), and extract the Levenshtein distance of the package name as part of the features. Finally, anomaly detection algorithm is used to detect malicious packages. 
In the development process of open source packages, developers will host the code in GitHub. 
When the code released in PyPI is inconsistent in Github, it means that the software package may be injected maliciously. To solve this problem, Vu et al.\cite{vu2021lastpymile} proposed a framework LastPyMile for identifying differences between software package construction artifacts and corresponding source code repositories, which can Monitor the security of registries such as PyPI.

\section{Conclusion}

In this paper, we constructed a dataset containing 4,669 malicious code and conducted a large-scale empirical study. The study revealed that malicious code in the PyPI ecosystem mainly exhibits a single function and is simpler than malicious code in other platforms. 
In addition, malicious code exhibits prevalent multiple malicious behaviours, with information stealing and command execution behaviours prominent. 
The attack strategies and triggering methods of malicious code vary across platforms, but they are highly adaptable and targeted and have the features of continuous evolution and mutation. 
Malicious code utilizes many anti-detection techniques to bypass detection. Finally, we found that the malicious code mainly enters end-user projects through source code installation and targets Linux systems as the primary target.

This empirical study provides valuable insights of the malicious code lifecycle in the PyPI ecosystem, which provides possible direction for future research. Moreover, these findings can serve as a reference for developing more effective security measures to mitigate the risk of malicious code attacks.

\section*{DATA AVAILABILITY}

The datasets of the studies can be publicly accessed at \href{https://github.com/lxyeternal/pypi_malregistry}{https://github.com/lxyeternal/pypi\_malregistry}.

\section*{Acknowledgment}

This research is supported by the National Research Foundation, Singapore, and the Cyber Security Agency under its National Cybersecurity R\&D Programme (NCRP25-P04-TAICeN). This work is also supported in part by the National Key Research and Development Program under Grant (2022YFB3305203), China, the National Natural Science Foundation of China under Grant (U2133208), and the Sichuan Youth Science and Technology Innovation Team under Grant (2022JDTD0014). Any opinions, findings and conclusions or recommendations expressed in this material are those of the authors and do not reflect the views of National Research Foundation, Singapore, the Cyber Security Agency of Singapore, the National Key Research and Development Program of China, the National Natural Science Foundation of China, and the Sichuan Youth Science and Technology Innovation Team.

\bibliographystyle{ieeetr}
\bibliography{ref}

\end{document}